\documentclass[twocolumn]{aastex62}

\graphicspath{{./}{figures/}}

\shorttitle{A Cool White Dwarf with Warm Dust}
\shortauthors{Debes et al.}
\begin{document}

\title{A 3 Gyr White Dwarf with Warm Dust Discovered via the Backyard Worlds: Planet 9 Citizen Science Project}

\correspondingauthor{John H. Debes}
\email{debes@stsci.edu}
\author[0000-0002-1783-8817]{John H. Debes}
\affiliation{Space Telescope Science Institute, 3700 San Martin Dr., Baltimore, MD 21218, USA}
\author{Melina Th\'evenot}
%\email{melina.t@web.de }
\affiliation{Backyard Worlds: Planet 9}
\author[0000-0002-2387-5489]{Marc J. Kuchner}
%\email{Marc.Kuchner@nasa.gov}
\affiliation{NASA Goddard Space Flight Center, Exoplanets and Stellar Astrophysics Laboratory, Code 667, Greenbelt, MD 20771}
\author{Adam J. Burgasser}
%\email{aburgasser@ucsd.edu}
\affiliation{Department of Physics, Center for Astrophysics and Space Sciences, Mail Code 0424, 9500 Gilman Drive, La Jolla, CA 92093-0424 USA}
\author[0000-0001-5106-1207]{Adam C. Schneider}
%\email{aschneid10@gmail.com}
\affiliation{School of Earth and Space Exploration, Arizona State University, Tempe, AZ, 85282, USA}
\author[0000-0002-1125-7384]{Aaron M. Meisner}
%\email{aaron.m.meisner@gmail.com}
\affiliation{National Optical Astronomy Observatory, 950 N. Cherry Ave., Tucson, AZ
85719, USA}
\altaffiliation{Hubble Fellow}
\author[0000-0002-2592-9612]{Jonathan Gagn\'e}
%\email{jonathan.gagne.1@gmail.com}
\affiliation{Institute  for  Research  on  Exoplanets, Université  de  Montréal, 2900  Boulevard  Édouard-MontpetitMontréal,  QC  Canada  H3T  1J4}
\author[0000-0001-6251-0573]{Jacqueline K. Faherty}
%\email{jfaherty17@gmail.com}
\affiliation{Department of Astrophysics, American Museum of Natural History, Central Park West at 79th St., New York, NY 10024, USA}

\author{Jon M. Rees}
%\email{jorees@ucsd.edu}
\affiliation{Department of Physics, Center for Astrophysics and Space Sciences, Mail Code 0424, 9500 Gilman Drive, La Jolla, CA 92093-0424 USA}

\author{Michaela Allen}
%\email{michaela.allen221@gmail.com}
\affiliation{Backyard Worlds: Planet 9}

\author[0000-0001-7896-5791]{Dan Caselden}
%\email{dancaselden@gmail.com}
\affiliation{Backyard Worlds: Planet 9}

\author{Michael Cushing}
\affiliation{Department of Physics \& Astronomy, The University of Toledo, Mail Stop 113, 2801 W. Bancroft St., Toledo, OH 43606, USA}

\author{John Wisniewski}
%\email{wisniewski@ou.edu}
\affiliation{Homer L. Dodge Department of Physics and Astronomy,
The University of Oklahoma,
440 W. Brooks St.,
Norman, OK 73019}

\author{Katelyn Allers}
%\email{katelynallers@gmail.com}
\affiliation{Physics and Astronomy Department,
Bucknell University,
701 Moore Ave, Lewisburg, PA 17837
Norman, OK 73019}

\author{The Backyard Worlds: Planet 9 Collaboration}
\affiliation{Backyard Worlds: Planet 9}
\author{The Disk Detective Collaboration}
\affiliation{Disk Detective}

%\collaborationName{The Backyard Worlds: Planet 9 Collaboration}
%\collaborationName{The Disk Detective Collaboration}

\begin{abstract}

Infrared excesses due to dusty disks have been observed orbiting white dwarfs with effective temperatures between 7200~K and 25000~K, suggesting that the rate of tidal disruption of minor bodies massive enough to create a coherent disk declines sharply beyond 1~Gyr after white dwarf formation. \replaced{We report the discovery}{We report the discovery that the candidate white dwarf LSPM~J0207+3331}, via the Backyard Worlds: Planet 9 citizen science project and Keck Observatory follow-up spectroscopy, \replaced{of a white dwarf}{is hydrogen-dominated} with a luminous compact disk (L$_{\rm IR}$/L$_{\star}$=14\%) and an effective temperature nearly 1000~K cooler than any known white dwarf with an infrared excess.  The discovery of this object places the latest time for large scale tidal disruption events to occur at $\sim$3~Gyr past the formation of the host white dwarf, making new demands of dynamical models for planetesimal perturbation and disruption around \replaced{white dwarfs}{post main sequence planetary systems}. Curiously, the mid-IR photometry of the disk cannot be fully explained by a geometrically thin, optically thick dust disk as seen for other dusty white dwarfs, but requires a second ring of dust near the white dwarf's Roche radius. In the process of confirming this discovery, we found that careful measurements of WISE source positions can reveal when infrared excesses for white dwarfs are co-moving with their hosts, helping distinguish them from confusion noise.
\end{abstract}

\keywords{white dwarfs --- planetary systems --- circumstellar matter --- stars:individual(LSPM~J0207+3331)}


\section{Introduction}
\label{sec:intro}
White dwarfs with infrared excesses hold critical clues to the long term evolution and chemistry of minor bodies in orbit around stars outside the Solar System. About 1-4\% of white dwarfs have detectable infrared excesses due to geometrically flat and optically thick dust disks \citep{vonhippel07, debes11,barber14,bonsor17}. These disks are believed to be caused by the tidal disruption of rocky bodies that then accrete onto the surface of the host white dwarf \citep{debes02,jura03}. Since white dwarfs are sensitive probes to the accretion of material, the atomic abundance of the dust is revealed through spectroscopy of the white dwarf photosphere \citep{zuckerman07}. Nearly 25-50\% of white dwarfs show dust accretion, implying that most white dwarfs have some flux of rocky bodies entering within their tidal disruption radius over their cooling lifetime \citep{koester14,hollands18}. While very old white dwarfs show photospheric evidence of dust accretion, dusty infrared excesses have previously only been seen around white dwarfs with T$_{\rm eff}>$7000~K or a cooling age of $\sim$1~Gyr \citep{farihi08,barber14}. 

We report the discovery that the high proper motion object \added{and candidate white darf}, LSPM~J0207+3331, \replaced{is a white dwarf}{has a hydrogen dominated atmosphere}  with an infrared excess. \added{\citet{LSPM} first identified LSPM~J0207+3331 as a high proper motion object, and \citet{gentile19} identified it as a candidate white dwarf (WDJ020733.81+333129.53) with T$_{\rm eff}$=5790 and $log~g$=8.06 if it had a hydrogen-dominated atmosphere.} \replaced{It}{We show through the construction of a multi-wavelength spectral energy distribution (SED) that it} possesses a T$_{\rm eff}$=6120~K and hosts a strong infrared excess with L$_{\rm IR}$/L$_{\star}\sim$0.13, consistent with a significant amount of dust filling the region interior to the Roche disruption radius of the white dwarf. 

The independent identification that this object was a white dwarf and the discovery of the infrared excess were made by citizen scientists of the Backyard Worlds: Planet 9 citizen science project.
This project is primarily designed to discover high proper motion WISE sources, with a focus on finding nearby brown dwarfs and new planets in the outer solar system \citep[][]{byw1}. The project's volunteers identify high proper motion sources by viewing flipbooks of four different epochs of unsmoothed WISE images \citep{meisner18} at a Zooniverse website (\url{www.backyardworlds.org}). They have also developed their own online tools for interrogating the WISE data \citep{2018ascl.soft06004C}.

In \S \ref{sec:discover} we discuss the discovery and confirmation of LSPM~J0207+3331 as a nearby white dwarf with an infrared excess. In \S \ref{sec:models} we derive the parameters for the host star and its dusty disk, and finally in \S \ref{sec:discuss} we discuss the significance of this discovery in the context of previously known dusty white dwarfs.

\section{Discovery and Confirmation of an Infrared Excess Around LSPM J0207+3331}\label{sec:discover}

LSPM~J0207+3331 was first flagged as an interesting candidate white dwarf by the Backyard Worlds: Planet 9 citizen scientist Melina Th\'evenot, who alerted the science team of its coordinates and its unusually red AllWISE $W1$-$W2$ color for a white dwarf: W1-W2=0.92$\pm$0.05, indicative of a significant IR excess consistent with a dusty disk \citep{hoard13}. Its AllWISE photometry is free from artifacts or obvious contaminants, with no significant variability in the Level 1b epoch data for either the WISE or NEOWISE photometry \citep{wright10}. Both photometry taken in 2010/2011 during WISE's main mission and over the 3.5 years of NEOWISE epochs were consistent within the uncertainties, and thus we average the $W1$/$W2$ magnitudes between the missions \citep{mainzer11,mainzer14}. We also obtained GALEX NUV photometry \citep{bianchi14}, $g,r,i,z$ photometry from the Pan-Starrs PS1 catalog \citep{chambers16}, and NIR photometry from 2MASS \citep{cutri03}. A table of all photometry obtained is listed in Table \ref{tab:t1}.

\begin{figure}
    \centering
    \plotone{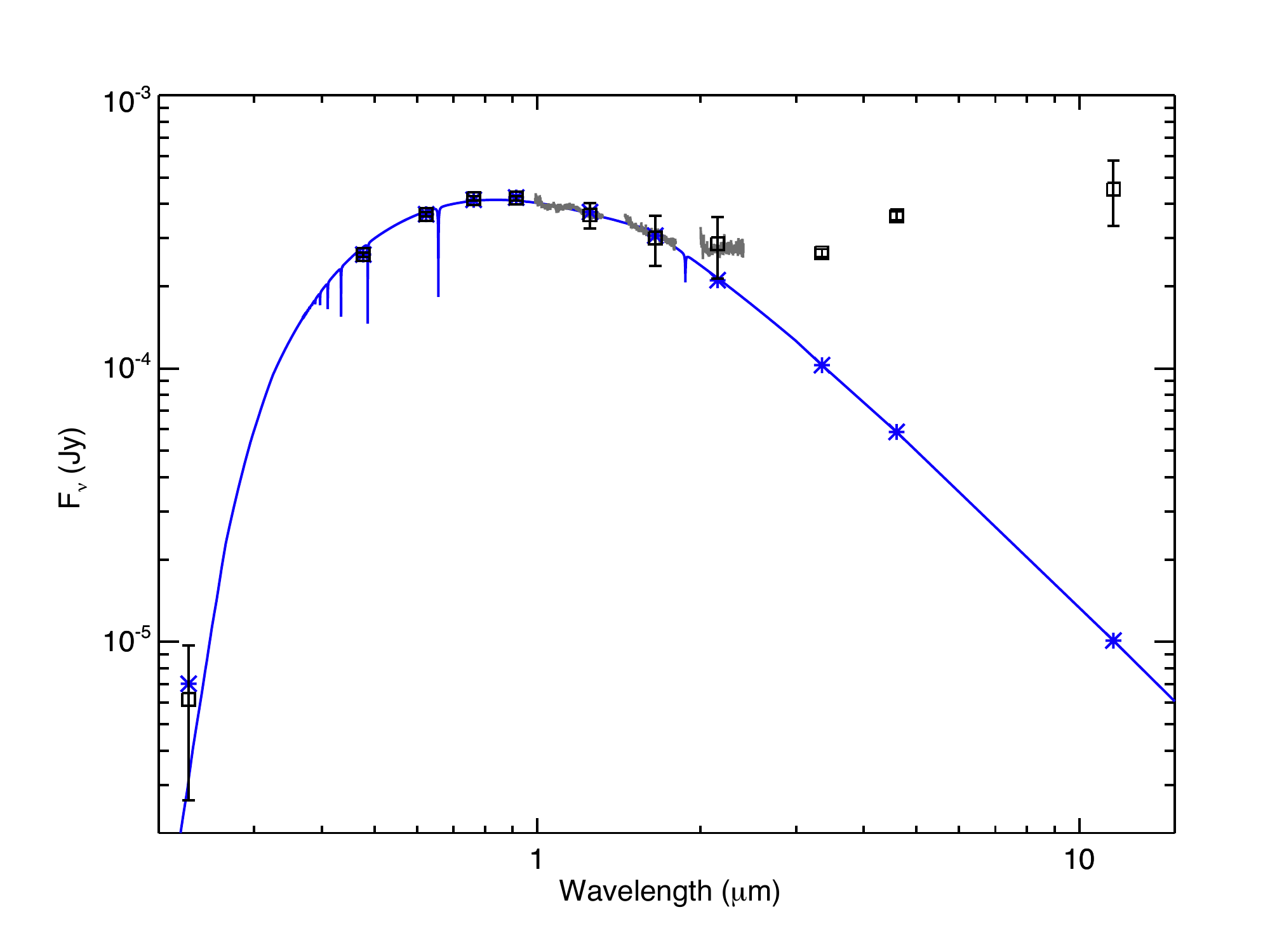}
    \caption{Spectral Energy Distribution of the LSPM~J0207+3331 system. Black squares represent GALEX NUV, Pan-STARRS $griz$, 2MASS, and AllWISE photometry. The gray lines represent the NIRES spectra. Overplotted are the predicted flux densities from our best fit white dwarf photometric model, and a comparable model from a publically available grid of DA spectra \citep{koester10,tremblay09}.}
    \label{fig:f1}
\end{figure}

\begin{deluxetable}{lc}
\tablecolumns{2}
\tablewidth{0pt}
\tablecaption{Photometry and Astrometry of LSPM~J0207+3331 \label{tab:t1}}
\tablehead{\colhead{Band} & \colhead{Mag} \\}
\startdata
GALEX NUV & 21.92$\pm$0.62 \\
\added{$G_{BP}$} & \added{17.87$\pm$0.01} \\
$g$ & 17.86$\pm$0.02 \\
\added{$G$} & \added{17.52$\pm$0.01}\\
$r$ & 17.49$\pm$0.02 \\
\added{$G_{RP}$} & \added{17.03$\pm$0.01} \\
$i$ & 17.34$\pm$0.02 \\
$z$ & 17.34$\pm$0.02 \\
$J_{\rm 2MASS}$ & 16.6$\pm$0.1 \\
$H_{\rm 2MASS}$ & 16.3$\pm$0.2 \\
$K_{\rm s,2MASS}$ & 15.9$\pm$0.3 \\
$W1$ & 15.08$\pm$0.03 \\
$W2$ & 14.20$\pm$0.03 \\ 
$W3$ & 12.2$\pm$0.3 \\
\hline
\added{Gaia DR2 ID} & \added{325899163483416704} \\
\added{GAIA DR2 Parallax} & \added{22.44$\pm$0.20~mas} \\
\added{GAIA DR2 $\mu_\alpha cos(\delta)$}  & \added{170.0$\pm$0.3~mas} \\ 
\added{GAIA DR2 $\mu_\delta$} & \added{-25.4$\pm$0.3~mas)} \\
\enddata
\end{deluxetable}
In additon to photometry, we obtained medium-resolution near-infrared spectra of LSPMJ0207+3331 using the Near-Infrared Echellette Spectrometer (NIRES) on the Keck II telescope \citep{wilson04}. The data were obtained on October 27th 2018 in partly cloudy conditions. NIRES has a fixed configuration, and covers a wavelength range of 0.94-2.45 $\mu$m at a resolution of $\sim$ 2700, using a slit width of 0\farcs55. 
We obtained 12 exposures of 300s each, nodding 5\arcsec\ along the slit between exposures for sky subtraction (3 complete ABBA sequences), at a mean airmass of 1.14 and with the slit aligned at the parallactic angle.
We also obtained spectra for HD13869, an A0V star with $V_{\mathrm{mag}}$ = 5.249, at a similar airmass for telluric correction and flux calibration. 
Dark and flat field frames were obtained at the end of the night. 
The data were reduced using a modified version of Spextool \citep{Cushing2004}, following the standard procedure. 
Wavelength calibration was determined using telluric lines, with a resulting RMS scatter of 0.073~\AA.
The spectra for each A-B pair were extracted individually and then combined together with the other extracted pairs.
The telluric correction procedure was carried out as detailed in \citet{Vacca2003}.

The final SED of the system is shown in Figure \ref{fig:f1} compared to a best-fit model of the white dwarf as discussed in \S \ref{sec:models}. As can be seen, a clear excess occurs just longwards of the $H$ band, confirmed both by the NIRES spectrum and the lower signal-to-noise ratio 2MASS photometry. The excess increases beyond 5$\micron$ with a low SNR detection of LSPM~J0207+3331 in $W3$, implying significant amounts of colder dust in the system. The inferred flux density at 12~$\micron$ is comparable to the peak flux density of the white dwarf.

The {\em Gaia} photometry and parallax of the system also implies that LSPM~J0207+3331 is a cool white dwarf \citep{gaia}. The parallax, 22.44$\pm$0.20~mas, implies a distance of 44.6$\pm$0.4~pc assuming a linear propagation of the parallax uncertainty. Based on {\em Gaia} photometry+parallax alone, LSPM~J0207+3331 is consistent with a mass of \replaced{$<$0.7~$M_\odot$ and a cooling age of $\sim$3~Gyr}{0.62~$M_\odot$ and a cooling age of $\sim$3~Gyr \citep{gentile19}}.

\begin{figure}
    \centering
    \plotone{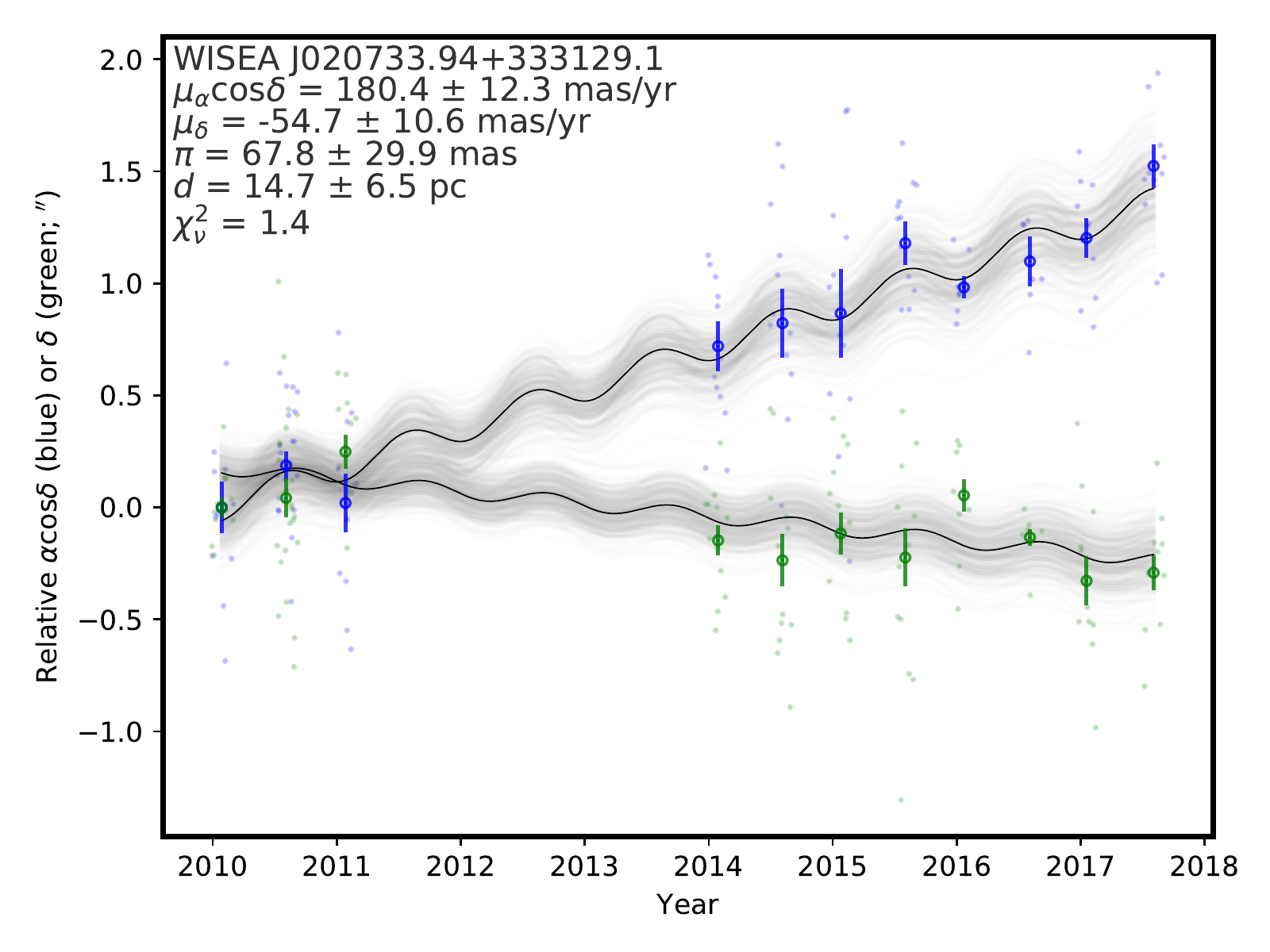}
    \caption{Proper motion fits to WISE and NEOWISE images of LSPM~J0207+3331, with blue symbols corresponding to proper motion in right ascension and green symbols corresponding to proper motion in declination. Our fits to the existing WISE detections of the white dwarf show that the mid-IR source has common proper motion to the {\em Gaia} visible detection. This common proper motion, which is consistent within the uncertainties, provides a direct confirmation that the infrared excess measured is co-located with the white dwarf and not background contamination. A parallax is not significantly detected with the WISE epochs alone.
    \label{fig:f2}}
\end{figure}

Following the initial recognition of its large proper motion in Backyard Worlds, we determined its WISE proper motion for WISE-only detections of LSPM~J0207+3331 via a Monte Carlo fitting \citep{theissen17} of the individual epoch positions, resulting in $\mu_\alpha cos(\delta)$=180$\pm$12~mas and $\mu_\delta$=-55$\pm$11~mas (See Figure \ref{fig:f2}). This is consistent within the uncertainties to the reported proper motion ($\mu_\alpha cos(\delta)$=170.0$\pm$0.3~mas and $\mu_\delta$=-25.4$\pm$0.3~mas) of the object in the {\em Gaia} DR2 catalog \citep{gaia} and the LSPM catalog \citep{LSPM}. 

 The combination of a common proper motion to the WISE mid-IR source and the presence of IR excess in the NIRES spectrum (with maximum angular extent of 1\farcs5), confirms that the excess is physically associated with LSPM~J0207+3331 and located $<$67~AU from the star.

\begin{figure}
    \centering
    \plotone{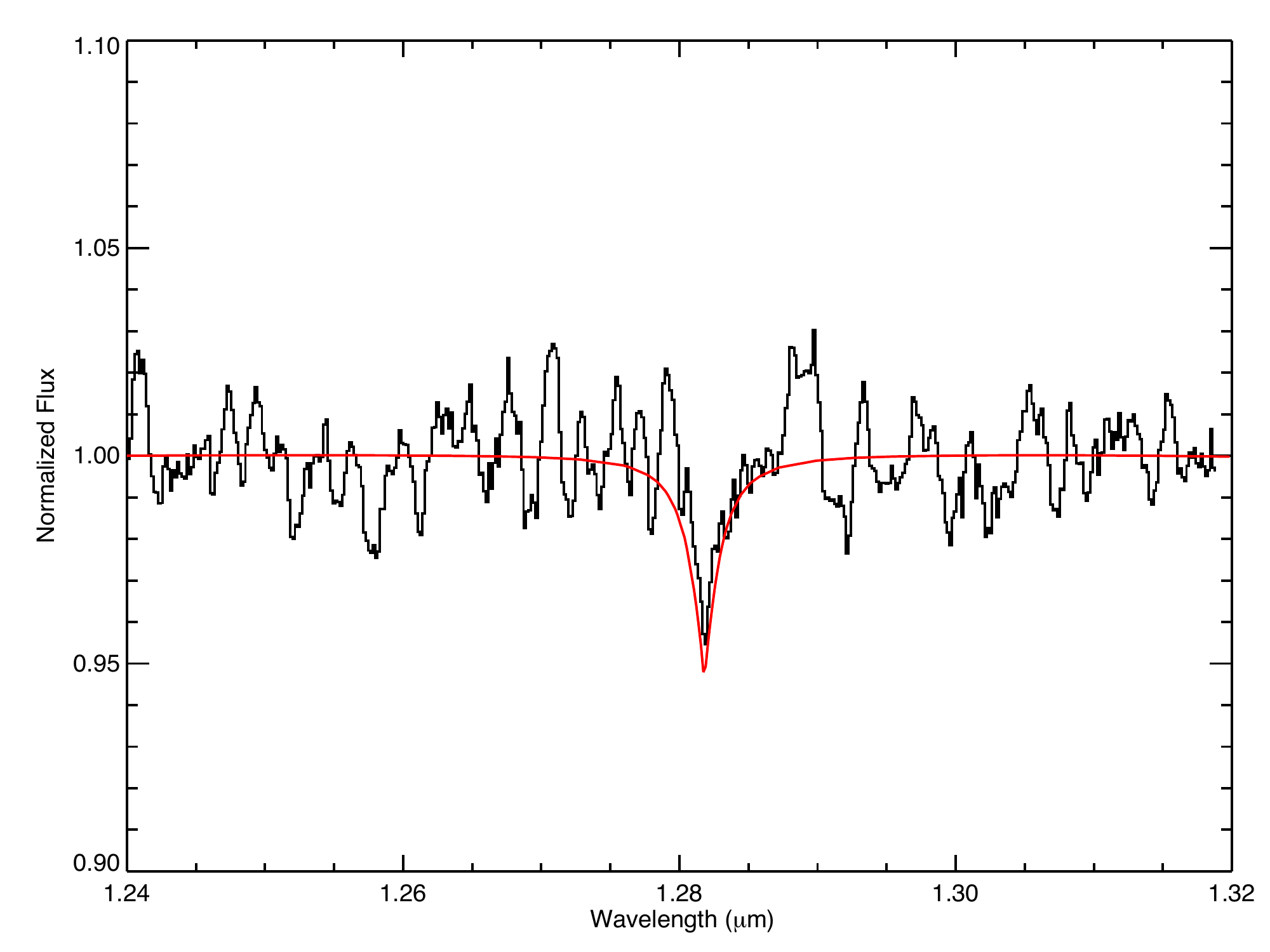}
    \caption{A 7-pixel boxcar smoothed NIRES spectrum of LSPM~J0207+3331 around the Pa-$\beta$ absorption line. The red line is an interpolated model from publicly available cubically interpolated DA synthetic spectra \citep{koester10,tremblay09} with T$_{\rm eff}$=6120~K and log~g=8.16. The line depth provides strong evidence that the white dwarf is a DA and is not suffering from any interstellar extinction which might produce a spuriously low T$_{\rm eff}$.}
    \label{fig:f3}
\end{figure}

\section{Determining the Origin of the Infrared Excess}
\label{sec:models}

Before modeling the infrared excess, we used \deleted{our} existing observations to constrain the atmospheric composition, effective temperature (T$_{\rm eff}$) and gravity of LSPM~J0207+3331. Figure \ref{fig:f3} shows a weak detection of the hydrogen Paschen $\beta$ line at 1.28$\micron$ \added{in the NIRES spectrum}, making LSPM~J0207+3331 a DA white dwarf. We then fit the NUV, visible, and $J$ photometry to DA white dwarf cooling models\footnote{http://www.astro.umontreal.ca/\~bergeron/CoolingModels/} to infer the white dwarf properties \citep{holberg06,kowalski06,tremblay11,bergeron11}. To obtain a best fit, we minimized a $\chi^2$ metric over our using the reported {\em Gaia} parallax. \replaced{From the 1-$\sigma$ uncertainties on the distance we determined a systematic uncertainty to our parameters in addition to statistical uncertainties determined by enforcing a 95\% confidence interval.}{We estimated statistical uncertainties within a 95\% confidence interval}. \deleted{The systematic uncertainties are less than the statistical uncertainties and are not reported.} The best-fit T$_{\rm eff}$=6120$^{+48}_{-57}$~K, while the best-fit gravity is 8.16$\pm0.03$. Based on the cooling models we used, this implies M=0.69$^{+0.01}_{-0.02}$~M$_\odot$ and a cooling age of 3$\pm$0.2~Gyr. Based on the best fit temperature, mass, and gravity, we infer a radius of 0.011~R$_\odot$, and a bolometric luminosity of 1.6$\times$10$^{-4}$ L$_\odot$. \added{Our temperature and gravity determinations are similar to those reported by \citet{gentile19}, although they report a lower temperature (T$_{\rm eff}$=5790$\pm$110~K) and lower gravity (8.06$\pm$0.07) than what we find using the Pan-STARRS photometry.}

\begin{figure*}
    \centering
    \plottwo{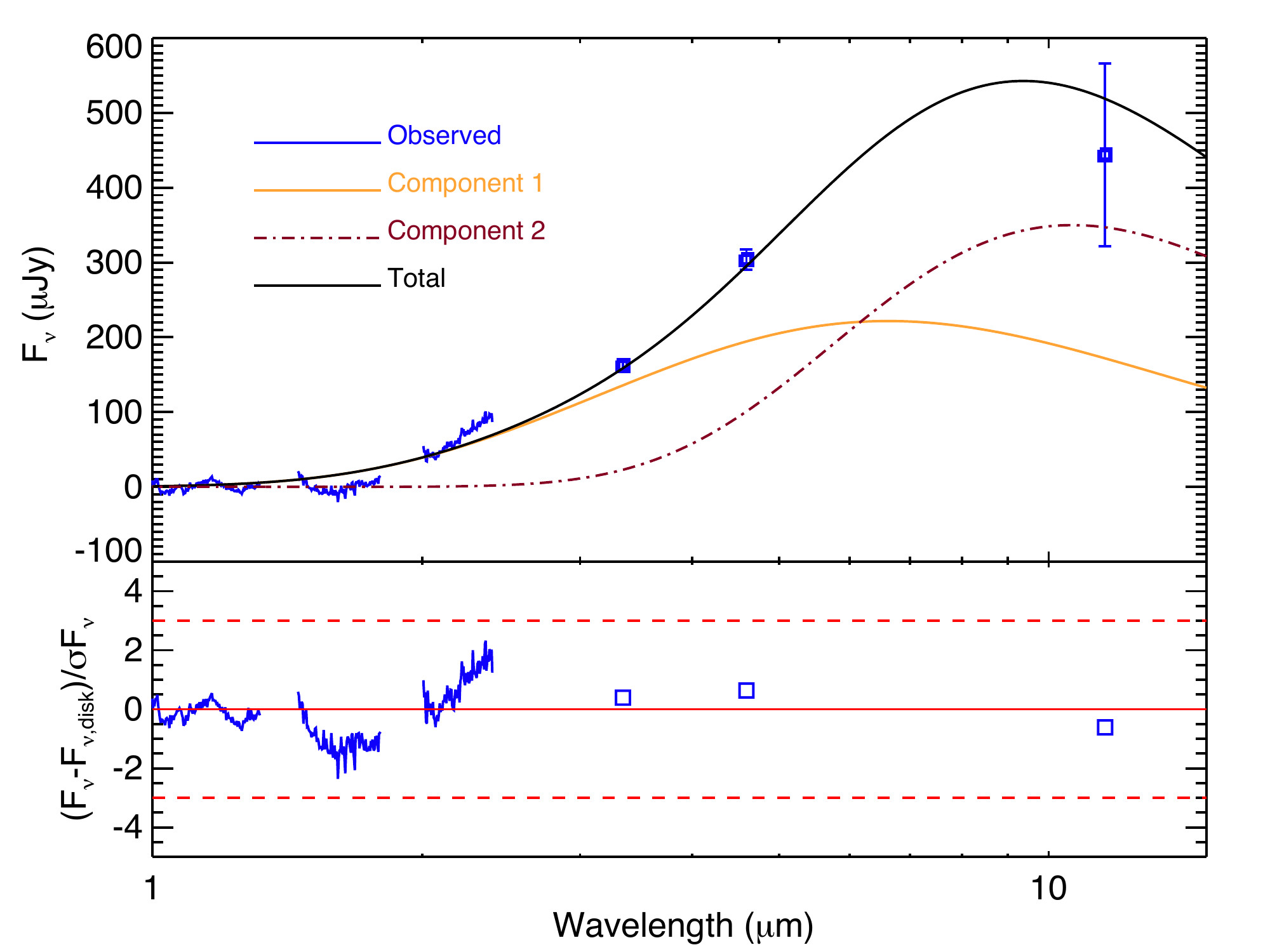}{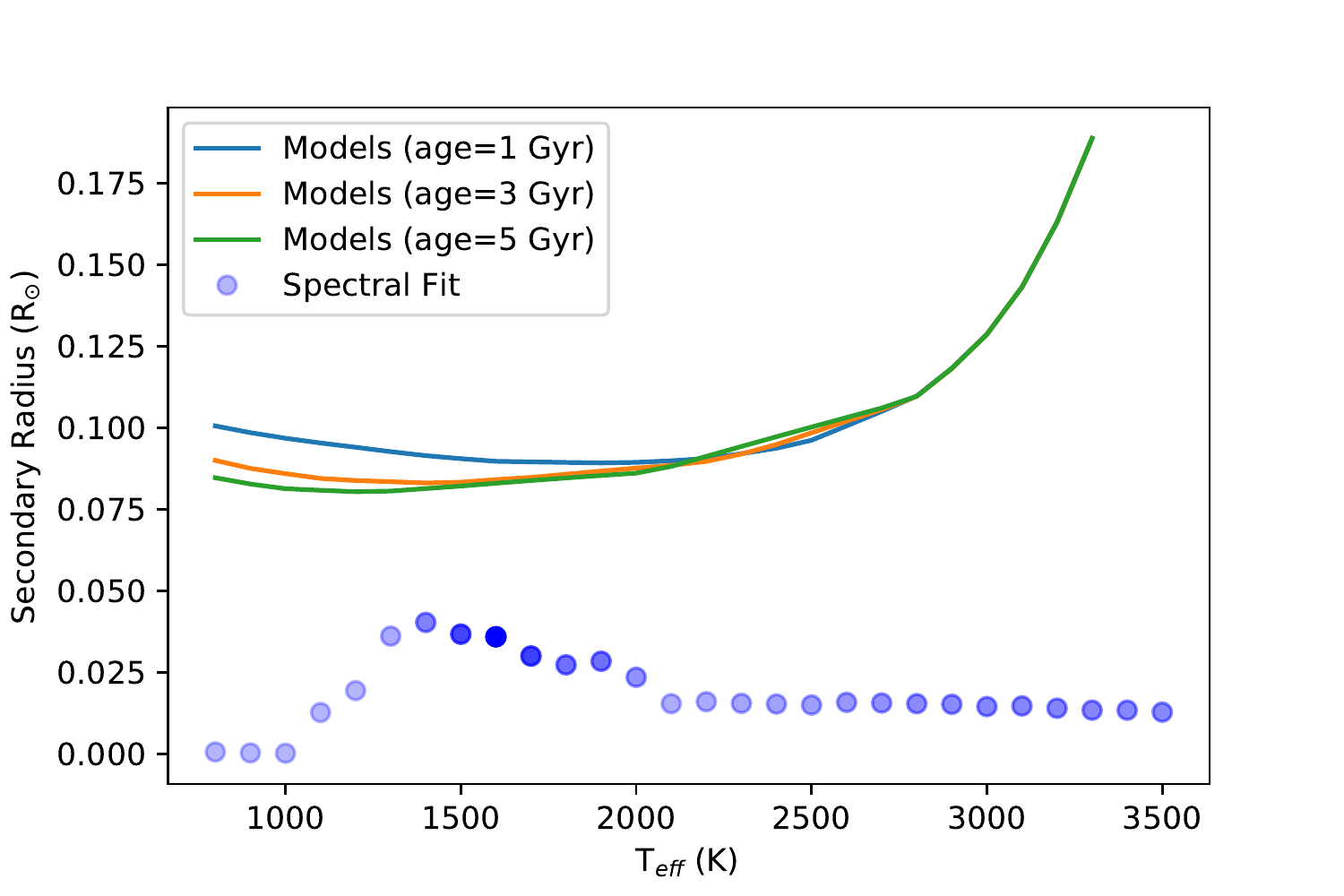}
    \caption{\added{(left) We investigated the possibility that the infrared excess was due to a cool companion by comparing different BT Settl models to the data. The best fit models implied anomalously small radii based on the NIRES spectrum. (right)}We subtracted off the flux from the best-fit white dwarf model to construct an SED of the residual infrared excess around LSPM~J0207+3331. Blue points are the rebinned NIRES spectrum and WISE photometry. We show a plausible model for the excess: a two-component disk that includes an optically thick inner region (Component 1; orange solid curve) and an optically thin structure near the edge of the WD tidal disruption radius (Component 2; purple dash-dot curve). Component 1 has an inner disk temperature of 1400~K and an outer temperature of 550~K. Component 2 is assumed to be an optically thin 480~K blackbody with an effective area of 10$^{21}$~cm$^2$. The lower panel shows the residuals against our disk model for the NIRES spectra and the WISE photometry, with the red dashed lines denoting $\pm$3$\sigma$ residuals.}
    \label{fig:f4}
\end{figure*}

\added{We first attempted to explain the excess as due to a cool dwarf companion. We compared the NIR spectrum to a series of binary models. We combined our spectral white dwarf models with companion spectra with temperatures ranging from 800~K to 3500~K, drawn from the BT Settl models \citep{2012RSPTA.370.2765A}. We assumed log~g = 5.0 below T$_{eff}$ = 1200~K and log~g = 5.5 above.  The secondary spectra were scaled to minimize the $\chi^2$ difference between observed and binary model spectra, and corresponding radii determined from this scale factor. For the lowest-temperature models (e.g., T dwarfs), the corresponding radii were vanishingly small ($<$ 0.01~R$_{\odot}$) as strong molecular features would otherwise contaminate the smooth continuum shape seen with NIRES. For the highest-temperature models (e.g., M dwarfs), radii were similarly small due to the higher surface fluxes. The best-fit model was for a T$_{eff}$ = 1600~K secondary, and while this was significantly better than a bare white dwarf photosphere, it still required a secondary radius (0.036~R$_{\odot}$) well below that predicted by evolutionary models  \citep[0.08~R$_{\odot}$][]{Burrows01,Baraffe03}. The left panel of Figure \ref{fig:f4} compares the best-fit secondary radii to evolutionary model predictions for 1, 3 and 5~Gyr; in all cases, the inferred radii are far below that expected for a hydrogen electron-degenerate object, ruling out the possibility that the excess arises from a companion.
}

We next attempted to fit the observed infrared excess with a face-on optically thick disk that fills the available orbital space as is traditional for most white dwarf disk modeling \citep{jura03,rocchetto15,dennihy17}. For LSPM~J0207+3331, this implies a maximum L$_{\rm IR}$/L$_\star$ of $\sim$0.13, assuming an inner disk temperature of 1800~K and an outer radius consistent with a Roche disruption radius of 0.69~R$_\odot$. This model failed to fit the observed data, overproducing flux between the NIR and $W1$ and underpredicting the observed flux at $W2$ and $W3$. Adjusting either the inner radius of the disk or the inclination forced a decent fit to the short wavelength data, but still failed to explain the fairly bright observed mid-IR flux beyond 4$\micron$. Similarly, increasing the outer radius of the disk to be consistent with a larger Roche radius as would be assumed for lower density material such as ice or porous rubble, did not solve the poor fit to the data.

As an alternative approach \added{to explain the observed SED}, we posit a hybrid combination of an optically thick face-on disk and a single temperature blackbody as a plausible fit to the observed photometry.  For the interior optically thick disk, we assume an inner temperature of 1400~K and outer temperature of 550~K, corresponding to an inner and outer radius of 0.047~R$_\odot$ and 0.21~R$_\odot$ respectively. We next assume a second component corresponding to a cooler optically thin dust distribution with a bulk temperature of 480~K. To fit the observed dust SED requires an effective surface area of 10$^{21}$~cm$^2$. The two components thus have a combined L$_{\rm IR}$/L$_{\star}$=0.14. The proposed model is compared to the observations in Figure \ref{fig:f4}, assuming that the NIRES spectra have a systematic 5\% uncertainty on the relative flux calibration. This model simultaneously fits the short wavelength emission as well as the longer wavelength emission. While background contamination in the mid-IR is possible, the common proper motion of the WISE $W1$ and $W2$ sources with LSPM~J0207+3331 suggests that contamination is unlikely. \replaced{Higher spatial resolution mid-IR imaging and}{Mid-IR} spectroscopy of this object with the James Webb Space Telescope should \replaced{fully resolve both the mystery of the dust SED and}{better constrain the structure and composition of the disk as well as} test whether there is a need for multiple components to the disk \added{as implied by the $W2$ and $W3$ photometry}.

 Another possible explanation for the large $W2$ and $W3$ fluxes that doesn't require cold dust remains. For highly magnetic white dwarfs, the accretion of material onto the white dwarf surface also results in infrared emission from cyclotron radiation \citep{debes12b,parsons13}. \added{However, the $W2$ single epoch photometry RMS over 94 epochs between 2014-2017 is 0.19~mag, while the median reported uncertainty on the photometry is 0.18~mag.} \added{One might expect large $W2$ variations for cyclotron radiation, so without further information we rate this alternative as unlikely.}

\added{Given the adopted temperature and effective surface area under the assumption of optically thin blackbody emission from a narrow ring of grains, we can speculate about the location and mass of the dust present.}\deleted{If we assume that there is an optically thin dust component with 480~K, we can infer both the likely structure of the dust and its location.} Assuming blackbody absorbing grains, we calculate that an optically thin distribution of dust will have an equilibrium temperature of 480~K at $\sim0.94~R_\odot$, conveniently near the Roche radius of the white dwarf assuming a bulk density of water ice, or just outside it if the dust is composed of refractory material with higher density. \added{The presence of this component over the $\sim$7~yr of the WISE mission means that Poynting-Robertson (PR) drag has not had time to remove grains and places a limit to the minimum grain radius of the dust distribution to $s\sim8$\micron\ for the mass and luminosity of LSPM~J0207+3331 \citep{gustafson}. If the dust is $\sim$8~\micron\ in radius, the total surface area ($A$) inferred by our blackbody fits implies a total minimum mass of $A/(\pi s^2)m_{\rm grain}$=3.4$\times10^{17}~g(\rho/1 g~cm^{-3})$ equivalent to a small asteroid or comet.}

\section{Discussion}
\label{sec:discuss}
The discovery of LSPM~J0207+3331 as a cool white dwarf with a signficant infrared excess makes new demands of models that seek to explain dust around white dwarfs. The first demand is that this discovery pushes the latest observable infrared excess to $\sim$3~Gyr after white dwarf formation, a factor of three larger than previous discoveries indicated. Current models for delivering planetesimals inside the tidal disruption radius of a white dwarf struggle to deliver the requisite mass of material to create an observable infrared excess beyond the first Gyr of a white dwarf's lifetime, especially if the white dwarf is not part of a binary.  \citep{bonsor10,debes12a,frewen14,veras14a, veras14b,veras15,veras17,mustill18,smallwood18}. 

Second, dynamical models must now reproduce the observed frequency of detectable disks as a function of cooling age out to 3~Gyr. We have queried the Gaia DR2 catalog along with AllWISE cross-matches for white dwarfs with similar $G-G_{RP}$ colors and distances $<$200~pc (implying a range of T$_{\rm eff}$ between 5000~K and 7000~K) with $W1$ and $W2$ detections to estimate the possible frequency of disks similar to LSPM~J0207+3331. The full ADQL constraints we used were:

\begin{eqnarray}
{\rm gaia.parallax>5} \\
{\rm AND~parallax/parallax\_error> 20} \\
{\rm AND~G_{abs} > 5(G-G_{RP})+10} \\
{\rm AND~G_{abs}<16.3} \\
{\rm AND~G_{abs}>9.5} \\
{\rm AND~G-G_{RP}>0.33} \\
{\rm AND~G-G_{RP}<0.62} \\
{\rm AND~W1<16}
\end{eqnarray}

We note that we do not apply many of the quality filters suggested to select white dwarfs by \citet{gentile19}, instead relying on high significance parallaxes $\pi/\sigma\pi>20$. Out of \replaced{397}{346} candidate white dwarfs, only \replaced{six}{five} had $W1-W2>0.3$ including LSPM~J0207+3331. The candidates are: LSPM~J1345+0504 (2 faint galaxies within 5\arcsec), 2MASS~J21403597+7739195 (WD+M), 2MASS~J18333593+5812176 (WD+M), and WISEA J032245.51+390445.0 (WD+M).  This \added{preliminary} census implies a dust disk frequency of $\sim$0.2\%, nearly an order of magnitude smaller than the frequency of dust disks around younger WDs.

Third, the infrared excess seen for this disk requires a second, colder ring of dusty material that could potentially signal the presence of a gap in the system, or a component of dust that extends beyond the outer edge of the inner disk. If the second ring is confirmed, it would be the first example of a two-component ring system around a dusty white dwarf. \replaced{If small disruptions are a common phenomenon, it may explain the hints of variability seen for other dusty white dwarfs in the mid-IR as new dust is created and eventually accreted over a few years via Poynting-Robertson drag. Such a structure is qualitatively consistent with ideas of multiple smaller disruptions feeding accretion onto white dwarfs \citep[i.e.][]{jura08}}{If the dust disk has a gap near 0.94~R$_\odot$, this implies the possibility of a body that continuously clears dust from the system, since the PR drag timescale is so short}.

Finally, the presence of an optically thick disk at late white dwarf cooling times affords an interesting test of dust disk accretion. The accretion rate from optically thick disks is likely mediated by PR drag and is thus dependent on the luminosity and temperature of the white dwarf \citep{rafikov11}. Models of disk accretion predict a slowly decreasing minimum accretion rate for a detectable disk, and older white dwarfs appear less likely to accrete at a high enough rate to yield an observable disk \citep{hollands18}. Based on the predictions of this model, LSPM~J0207+3331 should have an accretion rate larger than $\sim$3$\times$10$^{7}$~g~s$^{-1}$, a prediction that can be tested by taking an optical spectrum of this white dwarf to look for metal lines from accreted dust.

\acknowledgements
We would like to thank the
many Zooniverse volunteers who have participated in the Backyard Worlds: Planet 9 project. We would also like to thank the Zooniverse web development team for their work creating and maintaining the
Zooniverse platform and the Project Builder tools. Additionally we thank the anonymous referee for their suggestions in improving the manuscript. This
research was supported by NASA ADAP grant NNH17AE75I. 
Some of the data presented herein were obtained at the W. M. Keck Observatory, which is operated as a scientific partnership among the California Institute of Technology, the University of California and the National Aeronautics and Space Administration. The Observatory was made possible by the generous financial support of the W. M. Keck Foundation. The authors wish to recognize and acknowledge the very significant cultural role and reverence that the summit of Maunakea has always had within the indigenous Hawaiian community.  We are most fortunate to have the opportunity to conduct observations from this mountain.
This research has made use of the VizieR catalogue access tool, CDS, Strasbourg, France. The original description of the VizieR service was published in A\&AS 143, 23. This research has also made use of "Aladin sky atlas" developed at CDS, Strasbourg Observatory, France.  \added{DA spectroscopic models used in this paper were obtained from the Spanish Virtual Observatory Theoretical Model Services site.}
\facility{Keck:2 (NIRES)}
\facility{WISE}

\end{document}